\documentclass{article}
\usepackage[final]{infocog_neurips_2022}

\setcitestyle{square,numbers}

\usepackage[utf8]{inputenc} % allow utf-8 input
\usepackage[T1]{fontenc}    % use 8-bit T1 fonts
\usepackage{hyperref}       % hyperlinks
\usepackage{url}            % simple URL typesetting
\usepackage{booktabs}       % professional-quality tables
\usepackage{amsfonts}       % blackboard math symbols
\usepackage{nicefrac}       % compact symbols for 1/2, etc.
\usepackage{microtype}      % microtypography
\usepackage{xcolor}         % colors

\usepackage{graphicx}

\usepackage{amsmath} 

\usepackage{wrapfig}

\title{The Information Bottleneck Principle in Corporate Hierarchies}

\author{Cameron Gordon\\%\thanks{\texttt{cameron.gordon@adelaide.edu.au}}\\%
  Australian Institute for Machine Learning\\
  University of Adelaide\\
  \texttt{cameron.gordon@adelaide.edu.au} \\
}

\begin{document}

\maketitle

\begin{abstract}
%In recent years the Information Bottleneck Principle has emerged as a useful tool for understanding the structure and processing in deep neural networks. However, it has yet to be applied widely within economi

The hierarchical nature of corporate information processing is a topic of great interest in economic and management literature. Firms are characterised by a need to make complex decisions, often aggregating partial and uncertain information, which greatly exceeds the attention capacity of constituent individuals. However, the efficient transmission of these signals is still not fully understood. Recently, the information bottleneck principle has emerged as a powerful tool for understanding the transmission of relevant information through intermediate levels in a hierarchical structure. In this paper we note that the information bottleneck principle may similarly be applied directly to corporate hierarchies. In doing so we provide a bridge between organisation theory and that of rapidly expanding work in deep neural networks (DNNs), including the use of skip connections as a means of more efficient transmission of information in hierarchical organisations. 

\end{abstract}

\section{Introduction}
The corporate structure of firms has been the subject of economic enquiry since Coase (1937) \cite{coase_nature_1937,malmgren_information_1961}. In particular, the structures required for information processing and communication within firms has received much interest from economists, organisational theorists, institutional researchers, cyberneticists, and sociologists \cite{simon_administrative_1997,wiener_cybernetics_1948,beer_brain_1972,knapp_hierarchies_1989,williamson_markets_1973,markey-towler_i_2017,markey-towler_rules_2019,durugbo_modelling_2013}. Firms must acquire, process, and act upon information in order to effect a goal or strategic commercial objectives, often under conditions of ambiguity, partial observability, and uncertainty \cite{simon_administrative_1997,malmgren_information_1961}. This requires the coordination of many individuals whose bounded information capacity is substantially lower than that of an organisation as a whole \cite{simon_administrative_1997,gigerenzer_bounded_2001,wiener_cybernetics_1948,helbing_information_2006}. Employees are typically arranged in a hierarchy with lower levels reporting to higher and so on up until the board  \cite{simon_administrative_1997,simon_designing_1971,simon_architecture_1962}. While decision-making and delegation may occur at all levels of an organisation, a formal decision-maker (e.g. a board member or executive) may be separated by several layers of hierarchy from the original source of information relevant to their decision \cite{helbing_information_2006,simon_administrative_1997}. Indeed, as Simon (1997) notes "information most important to top managers comes mainly from external sources" \cite{simon_administrative_1997}. Market statistics, supply-chain information, changes to government policy, media reports, customer interactions, and sales figures feature among the external information relevant to decision-making. This aggregated set of information in most cases far exceeds the attention limit of any one individual to process; hence, the question of how to efficiently process, transmit, and filter decision-relevant information is of critical importance for firms \cite{helbing_information_2006,simon_administrative_1997,gigerenzer_bounded_2001}. 

The transfer of relevant information within a firm may be seen principally as a question of signal transmission. The well-developed theory of the \textit{information bottleneck principle} provides a means of analysing the communication of \textit{relevant information} that exists between two signals potentially involving multiple stages of processing \cite{tishby_information_1999}. Furthermore, this theory has recently emerged as a powerful tool for understanding the information processing that occurs in the similarly hierarchical structure of deep neural networks (DNNs) \cite{tishby_deep_2015,shwartz-ziv_opening_2017}. While the hierarchical similarity between DNNs and corporate hierarchies has been noted sporadically within the literature (e.g. \cite{marinescu_research_2015}), the information theoretic and organisational design consequences of this connection have received little examination.

The primary contribution of this paper is to provide a bridge between organisational theory and rapidly expanding work in deep neural networks and information bottleneck theory. We additionally explore a common component of deep neural networks (skip connections) and consequences for efficient corporate information processing. We view \citet{helbing_information_2006} as the prior literature most related to our own. The authors investigate information flow in hierarchical networks and highlight the importance of side-channels and shortcut connections for efficient information transmission, however they do not extend this to the information bottleneck principle or neural architectures.

\section{The Information Bottleneck Principle}

The information bottleneck principle introduced in \cite{tishby_information_1999} describes conditions for a compressed intermediate representation between two random variable signals (an input and an output), by reference to the mutual information shared between these signals and the representation \cite{tishby_deep_2015,tishby_information_1999}. The principle states that a compressed intermediate representation should contain the \textit{minimum} information with respect to the input while remaining predictive of the output  \cite{tishby_information_1999}. Consider two random variables $X$ and $Y$. The entropy of a random variable $X$ is a measure of its uncertainty \cite{cover_elements_2006}:

\begin{equation}
    H(X)=-\sum\limits_{x \in X}p(x)\log_2(p(x)).
\end{equation}

The mutual information between two random variables $I(X;Y)$ symmetrically describes the decrease in the uncertainty in $X$ that occurs given that $Y$ is known (and vice-versa) \cite{cover_elements_2006}: 

\begin{equation}
    I(X;Y)=H(X)-H(X|Y)=H(Y)-H(Y|X).
\end{equation}

Take an intermediate representation $\hat{X}$ between an input $X$ and output $Y$. Then $I(X;\hat{X})$ describes the information that is preserved between the input and the representation; and $I(Y;\hat{X})$ that between representation and the output. Using a control parameter $\beta$, the information bottleneck principle states that the compressed representation should minimise the Lagrangian \cite{tishby_deep_2015,tishby_information_1999,shwartz-ziv_opening_2017}: 

\begin{equation}
    \mathcal{L}[p(\hat{x}|x)] = I(X;\hat{X})-\beta I(Y;\hat{X}), 
\end{equation}

where $p(\hat{x}|x)$ is a mapping for all $x\in X$ satisfying the probability constraint $\sum\limits_{\hat{x}\in \hat{X}}p(\hat{x})=1$. The parameter $\beta$ controls the rate of compression and may be viewed as a constraint on transmission.

%The statistic satisfies the optimisation:
%$\underset{\theta}{\text{argmin }} I(X;\theta)-\beta I(Y;\theta)$$, where $\beta$ is a bottleneck control parameter for a chosen amount of compression and $\theta$ are parameters of the statistic. Simply stated, the information bottleneck principle describes an intermediate representation containing the \textit{minimal} information of the input, while still remaining predictive of the output, subject to the amount of compression. 

The information bottleneck principle has been applied widely to problems of information processing in decision science \cite{tishby_information_2011}, psychology \cite{zaslavsky_efficient_2018}, and prediction \cite{goldfeld_information_2020}. Importantly, it has been used to analyse information flows in deep feedforward neural networks \cite{tishby_deep_2015,shwartz-ziv_opening_2017,saxe_information_2019,goldfeld_information_2020}. A deep feedforward neural network involves a hierarchical processing of layers in which the output of each layer feeds directly to the next. Typically this is written as a composition of functions $f(x)=f_k(f_{k-1}(f\dots(f_1(x))))$ where $x$ is an input, $f_i(x)$ is a function applied at layer $i$, and $k$ is the number of layers \cite{goodfellow_deep_2016}.

\begin{figure}
    \centering
    \includegraphics[width=\textwidth]{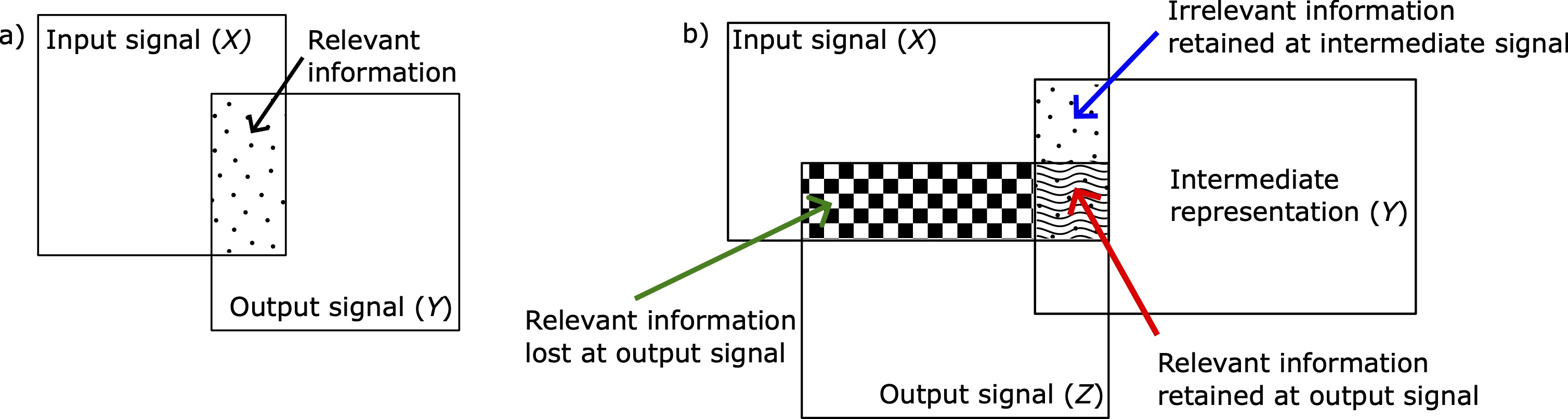}
    \caption{a) The relevant information is the minimal set of information in an input signal that is predictive of an output signal; without processing this corresponds to the mutual information $I(X;Y)$. b) Visualizing the data processing inequality. Taking an intermediate representation forms the Markov chain $X\to Y \to Z$. The mutual information between input signal $X$ and output signal $Z$ must be less than or equal to that between the input signal and the intermediate representation $Y$.}
    \label{fig:dpi}
\end{figure}

\citet{tishby_deep_2015} describe two properties that enable the information bottleneck principle to be applied to DNNs. The first is the data processing inequality, a fundamental result in information theory \cite{cover_elements_2006}. Briefly stated, for a Markov chain $X\to Y \to Z$ it must be that $I(X;Y)\geq I(X;Z)$. That is, that mutual information is non-increasing through processing. This may be seen visually by observing the sets of information shared between intermediate representations (see Figure \ref{fig:dpi}). Within a DNN information flow can be represented as a Markov chain $X \to T_1 \to T_2 \to \dots T_k \to Y$ where $T_i$ represents the $i^{th}$ hidden layer. The bottleneck representation implies that at convergence each layer will be the minimal representation required to predict the next layer subject to network capacity. Successive neural network layers hence reduce the mutual information shared between the input and output. The second property (invariance of mutual information to invertible transformations) enables the principle to hold for different re-parameterizations and intermediate structural forms \cite{zaslavsky_efficient_2018}. 

\section{Hierarchies in Corporations}
\vspace{-0.8pt}
Corporate organisations are typically arranged in a hierarchy, with reporting relationships from subordinates to superiors up to the board \cite{simon_administrative_1997,coase_nature_1937,dessein_organizations_2022,knapp_hierarchies_1989, malmgren_information_1961,foss_harald_1996}. Consider the following simplified model of a corporation with $N$ levels of hierarchy. For illustration, we first consider the simple case in which hierarchies are strict, and information flows only from one hierarchical layer to the next (i.e. information enters the corporation at layer 1 and is passed following processing to layer 2). At each layer $L_i$ some processing occurs (e.g. briefing notes, statistical analysis) with active decision taken by an individual on what information to transmit and what to filter from their superiors. We can note that we have the Markov chain $L_1 \to L_2 \to \dots \to L_N$, with the data processing inequality $I(L_1;L_2)\geq I(L_1;L_3) \geq \dots \geq I(L_1;L_N)$. Hence, this structure obeys a data processing inequality. By making the assumption that transformations within layers are invertible (i.e. prior to transmission the input information to that layer may be fully recovered), we can begin to apply the information bottleneck principle to this corporate hierarchy. 
\begin{wrapfigure}{r}{0.5\textwidth}

%\begin{figure}
    \centering
    \includegraphics[width=0.5\textwidth]{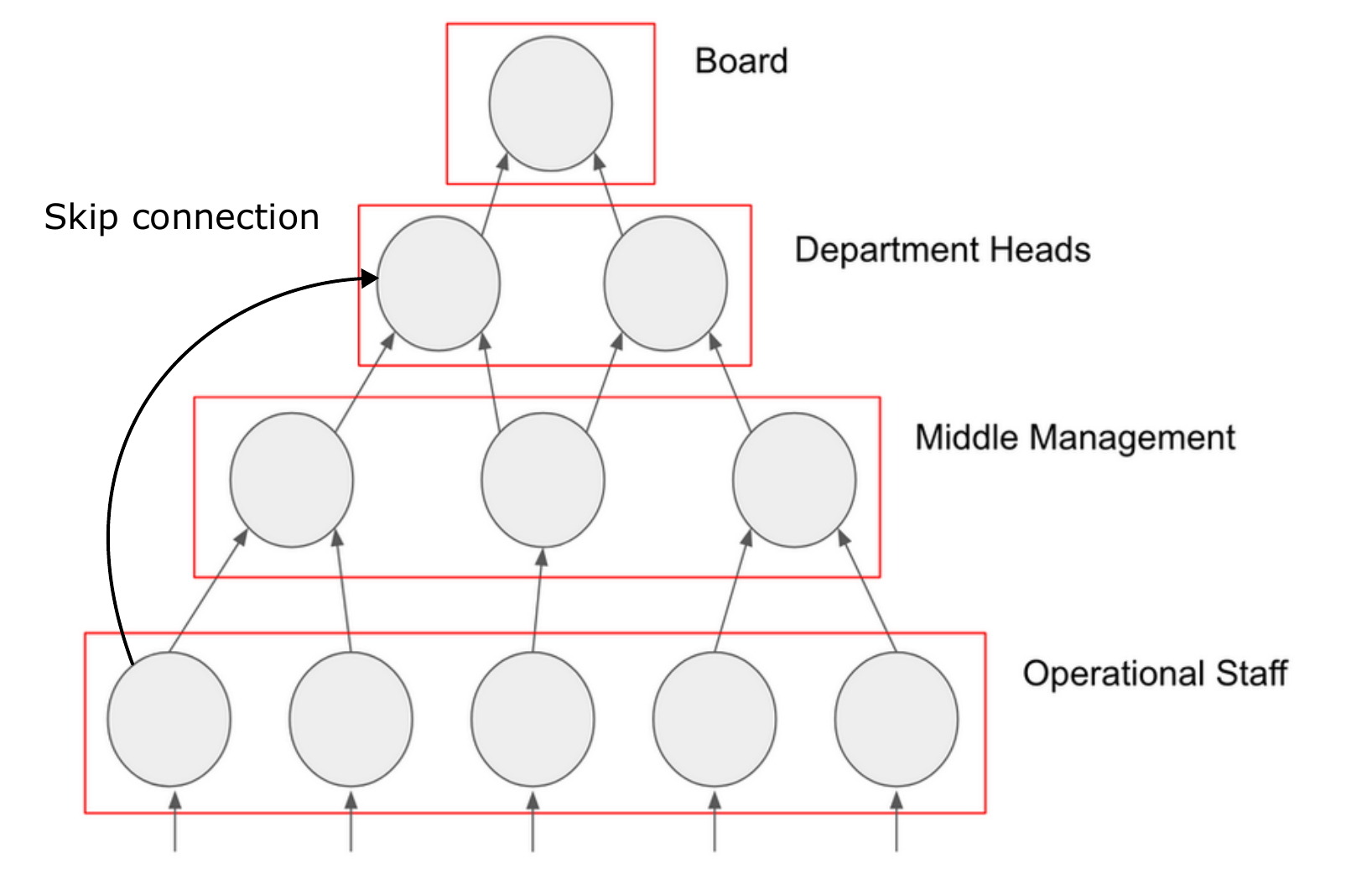}
    \caption{Information flow through a hierarchical organisation. Information entering an organisation at the operational level may have relevance to the board and is transmitted primarily through reporting lines. Skip connections within the organisation (informal and formal) enable lower level representations to be re-transmitted higher in the hierarchy, reducing the risk of incorrect filtering. }
    \label{fig:my_label}
    
%\end{figure}  
\end{wrapfigure}

%\subsection{Attentional bottlenecks}
A chief information constraint within firm management is the limited attention capacity of managers, who each operate within the constraints of bounded rationality \cite{simon_administrative_1997,gigerenzer_bounded_2001,helbing_information_2006}. Hence we may frame an information processing objective at each level in a hierarchy as transmitting the maximally relevant information subject to it being less than the attention capacity of managers at the next level. We can note this is equivalent to filtering or removing irrelevant information from being transmitted. Assume for the purposes of a decision that the attention constraint for a \textit{relevant decision-maker} at level $k$ is $\alpha_k$. With firm external information $X$ we have following optimisation at level $k-1$ of the organisation to produce a constrained information representation $L_{k}$ based on the preceding level: 

\begin{equation}
    \underset{L_{k-1}}{\text{argmin }} I(L_{k-2};L_{k-1})-\alpha_k I(L_{k-1};L_{k}), 
\end{equation}

which an be recognised as an information bottleneck. Similar analysis can be extended to lower levels in the hierarchy where the attention constraint for the decision-maker is replaced by the sum of attentions of individuals $j$ in the following layer: 

\begin{equation}
    \underset{L_{k-2}}{\text{argmin }} I(L_{k-3};L_{k-2})-\sum_{j} \alpha_{j,k-1} I(L_{k-2};L_{k-1}). 
\end{equation} 

And finally to the operational level where the signal $X$ enters the organisation: 

\begin{equation}
    \underset{L_{1}}{\text{argmin }} I(X;L_{1})-\sum_j \alpha_{j,1} I(L_{2};L_{1}). 
\end{equation} 

We can view the objective at each level as \textit{decoding} the information received and \textit{encoding} a compressed representation that can be transmitted through relevant reporting relationships \cite{tishby_deep_2015,shwartz-ziv_opening_2017}. 

Aligning the encoding-decoding process at each layer is required to reduce error in prediction and decision-making. This requires the use of top-down \textit{feedback}, a component known to be a critical mechanism for learning across domains \cite{wiener_cybernetics_1948,sutton_reinforcement_2018}. In deep neural networks feedback usually occurs through backpropagation of an error signal \cite{rumelhart_learning_1986,goodfellow_deep_2016}. In corporate structures feedback occurs through both formal and informal channels, including performance reviews, incentives, supervision, hiring decisions, and formal direction in order to align individual incentives to strategic goals. \cite{simon_administrative_1997,coase_nature_1937,williamson_markets_1973,malmgren_information_1961}. 

%\textit{Feedback} is transmitted in a top-down manner to better align the encoding-decoding process to the overall organizational goals.   

\section{Skip Connections in Corporate Hierarchies}
Much of the information relevant to corporate objectives is observed at the operational level and must be transmitted up the chain \cite{simon_administrative_1997}. In passing through several intermediate layers of a strict hierarchy decision-relevant information 
may be lost due to the data processing inequality, for example through misinterpretation or incorrect filtering by a manager. As \citet{simon_administrative_1997} writes, "A major communications problem, then, is that much of the information relevant to the decisions at this level originates at lower levels, and may not ever reach the higher levels unless the executive is extraordinarily alert." 

One resolution of this issue is that communication networks in corporations are typically \textit{not} strict hierarchies, but instead combine a mixture of formal and informal channels connecting multiple levels in the organisation \cite{simon_administrative_1997, helbing_information_2006}. It is well-known that different managers selectively attend to different details, and consultation with multiple groups occurs during making business decisions in practice \cite{cyert_observation_1956,dearborn_selective_1958,simon_administrative_1997}. The presence of shortcut, side-channel, and skip-connections has the effect of combining multiple filtered sources of an incoming signal, enabling error correction and multiple levels of compression to be viewed simultaneously \cite{helbing_information_2006}. An example of this is the transmission of a full report alongside a highly compressed executive summary. If relevant information has been incorrectly filtered by the summary it can be recovered through the full report. Similarly, lower-level analysts are often made present while their superiors are reporting to executives. The executive may rely primarily on the compressed representation presented by the superior as a matter of efficiency; and seek additional details, confirmation, or clarification from the subordinate only if required. 

The concatenation of a compressed representation with that of an earlier stage in a hierarchy can be recognised in deep learning as a form of skip connection, which have wide use in \textit{ResNets}, ResNeXTs, DenseNets, and related architectures, enabling extremely deep networks to be trained without the loss of signal and vanishing gradient known to affect deep strictly sequential networks \cite{he_deep_2015, he_identity_2016,xie_aggregated_2017,huang_densely_2018,pascanu_difficulty_2013,hochreiter_vanishing_1998,goodfellow_deep_2016}. Skip connections can also be viewed as \textit{bridges} which reduce path depth when viewed through the lens of network theory or communication topology \cite{momennejad_bridge_2019,momennejad_collective_2022,zhou_graph_2021}. %As \citet{helbing_information_2006} note the use of skip-connections or side-links may improve information redundancy and speeds information transfer through organisations.

%Known to influence collective intelligence, this architectural innovation may be similarly applied to models of corporate information processing. 

%In Appendix \ref{appendix} we present brief analysis for the flow of mutual information through skip connections demonstrating that a data processing inequality is satisfied by turning a $k$-order Markov chain into an equivalent $1$-order Markov chain for ResNets and DenseNets \cite{he_deep_2015,huang_densely_2018}. Related to this we may state that the mutual information retained through skip connections exceeds that of chain networks. 

%\begin{wrapfigure}{r}{0.5\textwidth}

%\end{wrapfigure}

%Residual connections \cite{he_deep_2015}

%As Herbert Simon noted in 1962, hierarchical architectures emerge when the complexity at a given level exceeds a given threshold \cite{simon_architecture_1962}.

\section{Conclusion} 
Corporations need to filter, compress, and transmit information internally in order effect decisions. In this paper we have argued that the information bottleneck principle may be used to analyse information processing within these hierarchical systems. In doing so we provide a bridge between work in organisational theory and recent advances in the study of neural networks. To an extent these systems face an analogous problem of making decisions under constrained resources: computational and financial \cite{coase_nature_1937,simon_administrative_1997,malmgren_information_1961,foss_harald_1996}. While we have pointed to the example of skip connections as an example of a topological structure in both systems it is an interesting open question whether other features known to improve neural network performance may play an analogous role in organisational design. 

\section*{Acknowledgements}
The author wishes to thank Dr. Lachlan MacDonald, Dr. Hemanth Saratchandran, Dr. Patrick O'Callaghan, and Mr. Keaton Jenner for their helpful suggestions and comments on draft forms of this paper, and Dr. Nan Ye for his help in understanding the details of information bottleneck theory.
\appendix

%\bibliographystyle{neurips_2022}
%\bibliography{references.bib}

\bibliographystyle{plainnat}
\bibliography{references.bib}

\end{document}